\documentclass[useAMS,usenatbib]{mnras}
\usepackage{graphicx}
\usepackage{dcolumn}
\usepackage{bm}

\usepackage{amsmath}
\usepackage{amssymb}

\newcommand{\rvir}{R_{vir}}
\newcommand{\mvir}{M_{vir}}
\newcommand{\vesc}{v_{esc}}

\newcommand{\dba}{\Delta m^2_{21}}
\newcommand{\dbc}{\Delta m^2_{23}}

\title[The relic neutrino composition as seen from Earth]{The relic neutrino composition as seen from Earth}

\author[A. N. Baushev]{A. N. Baushev$^{1}$\\
    $^{1}$Bogoliubov Laboratory of Theoretical Physics, Joint Institute for Nuclear Research\\
    141980 Dubna, Moscow Region, Russia}
\begin{document}

\date{}

\pagerange{\pageref{firstpage}--\pageref{lastpage}} \pubyear{2020}

\maketitle

\label{firstpage}

\begin{abstract}
Being generated, the relic neutrino background contained equal fractions of electron $\nu_e$, muon
$\nu_\mu$, and taon $\nu_\tau$ neutrinos. We show that the gravitational field of our Galaxy and
other nearby cosmic objects changes this composition near the Solar System, enriching it with the
heaviest neutrino $nu_3$. This mass state is almost free of the electron component (only $\sim 2\%$
of $\nu_e$) and contains more muon component than the tau one. As a result, the relic background
becomes enriched with taon and particularly muon neutrinos. The electron relic neutrinos are the
rarest for a terrestrial observer: instead of $1/3$, the relic background may contain only $\gtrsim
20\%$ of them.
\end{abstract}


\section{Introduction}
A very common cosmological consideration shows that, along with the relic radiation, there should
be relic neutrinos as well. Neutrinos were in the thermodynamical equilibrium with the other
substance in the early Universe and decoupled much earlier than the radiation did (when the
temperature of the Universe was $T_d\sim 2$~{MeV}). Soon after that the electron-positron pairs
annihilated, boosting the temperature of the photons, but not of the neutrinos. As a result, the
temperature of the neutrinos after that became $(4/11)^{1/3}\simeq 0.714$ of the photon one
\citep{gorbrub1}. Thus, the relic neutrinos occur in a very general cosmological scenario and their
number density is comparable with that of the relic photons.

There are three known types of neutrinos: electron $\nu_e$, muon $\nu_\mu$, and taon $\nu_\tau$
ones. The spin of all the neutrinos is $1/2$; however, only their state with left-handed helicity
seems to interact with anything else, and therefore only this state appears in any elementary
particle interaction. Modern theoretical models suggest that there can be more exotic neutrino-like
particles. For instance, heavy sterile neutrinos may form the main part of the dark matter.
However, we will consider only the three above-mentioned Standard Model neutrinos in this paper,
and we will hereafter imply only them by the term 'neutrino'.

Neutrino oscillations show that all the neutrinos have non-zero mass, and that $\nu_e$, $\nu_\mu$,
$\nu_\tau$ are not the eigenstates of the hamiltonian, and therefore they have no certain mass. We
denote the three mass states of neutrinos by $\nu_1$, $\nu_2$, $\nu_3$, and their masses by $m_1$,
$m_2$, $m_3$. The oscillations allow to measure the neutrino mass squared differences \cite[table
14.1]{pdg18}:
\begin{equation}
\dba\simeq 7.37\cdot 10^{-5}\; \text{eV}^2,\qquad \dbc\simeq 2.54\cdot 10^{-3}\; \text{eV}^2
 \label{23a1}
\end{equation}
It means that there should always be at least one neutrino with
$m\gtrsim(\dbc)^{1/2}+(\dba)^{1/2}\simeq 0.06$~{eV}.

The idea that cosmology sets a limit on the neutrino sum mass was offered by \citet{zel1966}.
Recent data on the Baryon Acoustic Oscillations \citep[eqn. (63b)]{planck2018} say that
\begin{equation}
m_1+m_2+m_3<0.120\; \text{eV}\qquad \text{95\% CL.}
 \label{23a2}
\end{equation}
Unfortunately, we do not know the masses of $\nu_1$, $\nu_2$, $\nu_3$: equations
(\ref{23a1}-\ref{23a2}) set only constraints on $m_1$, $m_2$, $m_3$. Historically, three limiting
scenarios were considered \cite[p. 254]{pdg18}
\begin{itemize}
 \item {\it Normal Hierarchical (NH):} $m_1< m_2 \ll m_3$,\quad $m_2\cong (\dba)^{1/2}\cong
 0.0086\;
 \text{eV}$,\quad $m_3\cong |\Delta m^2_{31}|^{1/2}\cong 0.0506\; \text{eV}$
 \item {\it Inverted Hierarchical:} $m_3\ll m_1 < m_2$,\quad $m_1\cong (|\dbc|-\dba)^{1/2}\cong 0.0497\;
 \text{eV}$,\quad $m_2\cong |\dbc|^{1/2}\cong 0.0504\; \text{eV}$
 \item {\it Most Degenerated:} $m_1\simeq m_2\simeq 0.05$~{eV}, $m_3\simeq 0.07$~{eV}
\end{itemize}
From the experimental point of view, the most degenerated scenario is now excluded, and the
inverted hierarchical one is highly unlikely \citep{neutrino2019}. We will consider only the normal
hierarchical scenario in this paper, though the final result can be trivially generalized for the
inverted hierarchical case. We will make a short comment about that in the end of the letter.

It is curious that, though the Universe is much more transparent for the relic neutrinos than for
relic photons, the relic photons may come to us from larger distances: if a neutrino is heavier
than $2\cdot 10^{-4}$~{eV}, it moves significantly slower than light and passes shorter distance
since the Big Bang \citep{b1983}.

Though the neutrino masses has not measured yet, recent experiments allow us to estimate them and
other neutrino parameters rather reliably. The values that we will use in this paper are not very
precise, since the paper is estimative in general. If the hierarchy is normal, $\nu_3$ is the
heaviest neutrino, and $\nu_1$ is the lightest one. From~(\ref{23a1}) and~(\ref{23a2}) we obtain
$0.08\; \text{eV}>m_3>0.06\; \text{eV}$, $0.02\; \text{eV}>m_2>0.0086\; \text{eV}$, $0.02\;
\text{eV}>m_1>0$. All the neutrinos, $\nu_1$, $\nu_2$, and $\nu_3$, are mixtures of $\nu_e$,
$\nu_\mu$, and $\nu_\tau$. Here we skip the details of calculation (which can be found, for
instance, in \citet{pdg18}), presenting only the final result: $\nu_1$ contains $\sim 67\%$ of
$\nu_e$, $\sim 8\%$ of $\nu_\mu$, and $\sim 24\%$ of $\nu_\tau$; $\nu_2$ contains $\sim 30\%$ of
$\nu_e$, $\sim 35\%$ of $\nu_\mu$, and $\sim 35\%$ of $\nu_\tau$; $\nu_3$ contains $\sim 2\%$ of
$\nu_e$, $\sim 57\%$ of $\nu_\mu$, and $\sim 41\%$ of $\nu_\tau$. The fact that the heavy neutrino,
$\nu_3$, contains only $\sim 2\%$ of the electron neutrino is the most important for us.

\section{The spectrum and density of the relic neutrinos in the absence of large-scale
structure in the Universe}
First of all, we need to calculate the spectrum and density of the relic
neutrinos in the absence of structures in the Universe. Then the Universe evolution can be
described by the Friedmann metric $ds^2=c^2 dt^2 - a^2(t) dr^2$, where $a(t)$ is the scale factor.
We accept that $a=1$ at the moment of the neutrino decoupling (we may do it, since the Universe is
flat).

Let us consider one type of neutrinos (or antineutrinos) of mass $m$. When they were in the
equilibrium in the early Universe, their number $dN$ in a volume $W$ with the absolute momentum
between $p$ and $p+dp$ was \citep{ll5}
\begin{equation}
dN=\dfrac{W p^2 dp}{2\pi^2 \hbar^3 (\exp[(\sqrt{m^2 c^4 +p^2 c^3}-\varpi)/T]+1)},
 \label{23a5}
\end{equation}
where $\varpi$ is the chemical potential, $T$ is the temperature. This equation can be
significantly simplified, since $mc^2\ll T$, $\varpi\ll T$ in the early Universe \citep{gorbrub1}.
If we denote the particle momentum at the decoupling temperature $T_d$ by $p_d$, we obtain for the
decoupling moment:
\begin{equation}
dN=\dfrac{W_d p_d^2 dp_d}{2\pi^2 \hbar^3 (\exp(p_d c/T_d)+1)}.
 \label{23a6}
\end{equation}
After the decoupling the neutrinos propagate freely in the Universe. The only two changes
experienced by the neutrinos are the red shift and number density decreasing as a result of the
Universe expansion. Consequently, in order to obtain the neutrino distribution at any moment of
time (or at any $a(t)$), we need to substitute $W_d=W/a^{3}(t)$, $p_d=p\cdot a(t)$ into
(\ref{23a6}):
\begin{equation}
dN=\dfrac{W p^2 dp}{2\pi^2 \hbar^3 (\exp(p c/T_\nu)+1)},\quad \text{where}\quad T_\nu\equiv
\dfrac{T_d}{a(t)}.
 \label{23a7}
\end{equation}
This equation would describe the relic neutrino distribution in the present-day Universe, if it was
perfectly uniform. Formally, it coincides with the thermal distribution of ultrarelativistic
fermions with $T=T_\nu$. However, though we will name $T_\nu$ 'the neutrino temperature', it is
important to understand that actually $T_\nu$ is no temperature. For instance, some of neutrino
types are now non-relativistic (at least, one of them has $m>0.05$~{eV}, while $T_\nu\simeq
2$~{K}~$\simeq 2\cdot 10^{-4}$~{eV} now), and distribution (\ref{23a7}) is non-equilibrium for this
sort of particles. There is no reason to be surprised: after the decoupling at $T=T_d$ the
neutrinos are perfectly collision-less and need not have any real thermodynamical temperature,
being out of thermodynamical equilibrium with anything.

It is best to determine $T_\nu$ from the well-measured temperature of the relic photons
$T_\gamma\simeq 2.7255$~K \citep{pdg18}. Then the present-day value $\tau$ of $T_\nu$ is $\tau=
(4/11)^{1/3} \: T_\gamma \simeq 1.945$~K~$\simeq 1.677\cdot 10^{-4}$~{eV}, and $T_\nu$ at a red
shift $z$ is equal to:
\begin{equation}
T_\nu(z)=(z+1)\tau.
 \label{23a8}
\end{equation}
Since $\tau^2$ is much smaller than $\dba$ and $\dbc$, at least two (or, possibly, all three)
neutrinos are now non-relativistic. The case of a relativistic neutrino is rather trivial: its
distribution cannot be significantly influenced by gravitational fields of the large-scale
structure, and it is still adequately described by (\ref{23a7}) and (\ref{23a8}). Let us consider
the case of non-relativistic neutrinos of mass $m$. Then $p=mv$, and we may rewrite (\ref{23a7})
as:
\begin{equation}
f(v)=\dfrac{m^3}{h^3 (\exp(v/\xi)+1)},\quad \text{where}\quad \xi\equiv c\cdot \dfrac{(1+z)\tau}{m
c^2}.
 \label{23a9}
\end{equation}
Here we introduce the distribution function $f$ in the 6-dimensional phase space, i.e. the number
of the neutrinos of a certain flavor in a 6-dimensional volume $dx^3 dv^3$ is $f dx^3 dv^3$. The
velocity parameter $\xi$ defines the characteristic speed of the relic neutrinos: the
root-mean-square speed is $\simeq 3.60\xi$. Substituting the mass ranges of the neutrinos $\nu_1$,
$\nu_2$, and $\nu_3$, we find that at the present epoch ($z=0$) the most massive neutrino $\nu_3$
has $630\;\text{km/s}<\xi< 840\;\text{km/s}$, and $\xi>2500\;\text{km/s}$ for $\nu_1$ and $\nu_2$.

\begin{figure}
    \resizebox{1.0\hsize}{!}{\includegraphics{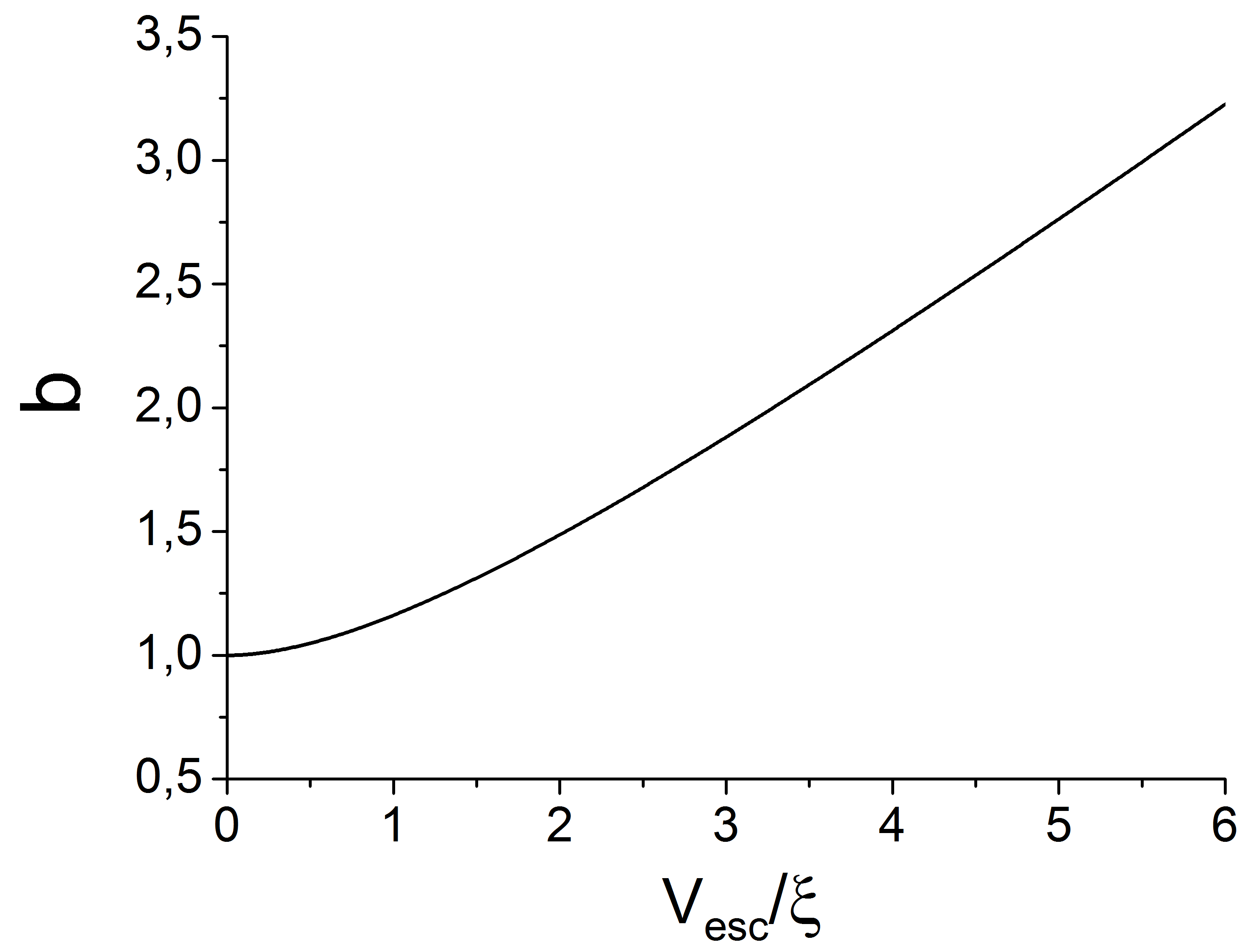}}
    \vspace{-0.0cm}
\caption{The dependence of the boost factor $b$ (see equation~(\ref{23a15})) from the ratio
$\sqrt{2\phi}/\xi=\vesc/\xi$ between the escape speed and the characteristic speed $\xi$ of the
neutrinos.} \label{23fig1}
\end{figure}

\section{The relic neutrino separation by space structures}
\subsection{The Galaxy} Distribution~(\ref{23a9}) is valid only in an absolutely homogenous
universe: the structures of the real Universe disturb function $f$ by their gravitational field.
Let us start from an estimation of the influence of our Galaxy. At the epoch of the Milky Way
formation ($z\simeq 4$) $\xi>3000\;\text{km/s}$ even for $\nu_3$. The capturing speed of the
protogalaxy can be estimated as $v_g^2/2=\sqrt{G\mvir/\rvir}$, where $\mvir\sim 10^{12} M_\odot$,
$\rvir\sim 200$~{kpc} are the Galaxy mass and radius. We obtain $v_g\simeq 200$~{km/s}, and
integrating distribution~(\ref{23a9}) we find that only $\sim 2.7\cdot 10^{-3} \%$ of neutrinos can
be captured, if $\xi=3000\;\text{km/s}$. Even taking into account that the neutrino density during
the Galaxy formation was $(z+1)^3\simeq 125$ times higher than the present-day value, we obtain
that the number of neutrinos $\nu_3$ captured inside the Galaxy is $\sim 300$ times lower than the
background given by~(\ref{23a9}). For the lighter neutrinos, $\nu_1$ and $\nu_2$, the captured
fraction is absolutely negligible.

However, the gravitational field of the formed Galaxy attracts neutrinos and increases their
density. Let us estimate the efficiency of this process. For simplicity, we consider a toy (but
rather close to realistic) model of our Galaxy as a single, stationary, spherically-symmetric halo
immersed into the field of neutrinos with distribution~(\ref{23a9}). We choose two concentric
spheres of radii $r_1$ and $r_2$ around the Galaxy center. The radius $r_1=8$~{kpc}, i.e., $r_1$ is
equal to the radius $r_\odot$ of the Sun orbit around the Galaxy center. We choose $r_2\gg \rvir$,
i.e., it is so large that we may neglect the influence of the gravitational field of the Galaxy at
$r_2$. We denote the distribution function $f$, the angle $\theta$ between the direction to the
center and the particle trajectory, and the particle speed at $r_1$ and $r_2$ by $f_1$, $\theta_1$,
$v_1$ and $f_2$, $\theta_2$, $v_2$, respectively.

Let us consider the number of neutrinos $f_1 dx^3 dv^3$ crossing the sphere $r_1$ in an angle
interval $[\theta_1; \theta_1+d\theta_1]$ in a speed interval $[v_1; v_1+dv_1]$ in a time interval
$dt$. They occupy the volume $dx^3= 4\pi r_1^2\cdot v_1\cos\theta_1 dt$ and the velocity volume
$dv^3=2\pi v_1^2 d(-\cos\theta_1) d v_1$. The neutrinos cross the sphere $r_2$ in an angle interval
$[\theta_2; \theta_2+d\theta_2]$ in a speed interval $[v_2; v_2+dv_2]$. Since the system is
stationary, they cross $r_2$ in the same time interval $dt$. We obtain:
\begin{eqnarray}
\nonumber &f_1\cdot 2\pi v_1^2 d(-\cos\theta_1) d v_1\cdot 4\pi r_1^2\cdot v_1\cos\theta_1 dt=\\
&=f_2\cdot 2\pi v_2^2 d(-\cos\theta_2) d v_2\cdot 4\pi r_2^2\cdot v_2\cos\theta_2 dt.
 \label{23a10}
\end{eqnarray}
One may easily simplify this equation to:
\begin{equation}
f_1 r_1^2 v_1^3 d\sin^2\theta_1 dv_1=f_2 r_2^2 v_2^3 d\sin^2\theta_2 dv_2.
 \label{23a11}
\end{equation}
Describing the neutrino motion, we may use the energy and angular momentum conservation:
\begin{eqnarray}
 \label{23a12} v_1^2&=& v_2^2+2\phi,\\
r_1 v_1\sin\theta_1&=& r_2 v_2\sin\theta_2,
 \label{23a13}
\end{eqnarray}
where $\phi$ is the gravitational potential at $r_1=8$~{kpc}, at the Sun orbit. From~(\ref{23a12})
we obtain $v_1 dv_1=v_2 dv_2$. Substituting it and~(\ref{23a13}) to~(\ref{23a11}), we find that
$f_1=f_2$.

Of course, this result is predictable: it is just a consequence of the Liouville's theorem, the
phase-space distribution function is constant along the trajectories of the particles. However, the
derivation directly demonstrates that the distribution function near the Earth is isotropic.
Moreover, since $r_2\gg r_1$, only particles with $\theta_2 \simeq 0$ may reach $r_1$. Thus, even
if the distribution function $f_2$ far from the center is strongly anisotropic, the distribution
near Earth $f_1$ is most likely isotropic: it is sufficient to have $f_2(\theta_2=0)=\it{const}$ at
the sphere $r_2$. The isotropization of the distribution function of collisionless particles
towards the center of a spherical halo is a general property \citep{14}.

For convenience, we denote the distribution function of the neutrinos at the Sun orbit in the
Galaxy (i.e., $f_1$) by $F$, and the neutrino velocity at this radius (i.e., $v_1$) by $V$. Since
$f_1=f_2$, we need just to substitute~(\ref{23a12}) into~(\ref{23a9}) to obtain $F$:
\begin{eqnarray}
\nonumber F(V)&=&\dfrac{m^3}{h^3 (\exp(\frac{\sqrt{V^2-2\phi}}{\xi})+1)},\quad \text{for}\quad
V\ge\sqrt{2\phi}\\
F(V)&=&0,\quad \text{if}\quad V<\sqrt{2\phi}.
 \label{23a14}
\end{eqnarray}
Instead of the gravitational potential $\phi$, one may use the escape speed $\vesc\equiv\sqrt{2
\phi}$. The density of neutrinos corresponding to the undisturbed distribution $f$ is $n=\int
f(\vec v) dv^3$, the density of neutrinos near the Earth is $n_\odot=\int F(\vec V) dV^3$. We may
introduce the boost factor $b\equiv n_\odot/n$, i.e.,
\begin{equation}
b=\int^\infty_{\vesc}\! F(\vec V) dV^3\left/\int^\infty_0\; f(\vec v) dv^3\right..
 \label{23a15}
\end{equation}
The boosting factor $b$ depends only on the ratio $\sqrt{2\phi}/\xi=\vesc/\xi$, i.e., the ratio
between the escape speed and the characteristic speed $\xi$ of the neutrinos. The dependence is
presented in figure~\ref{23fig1}.

The physical mechanism of the boosting is apparent: the gravitational field of the Galaxy
accelerates the neutrinos, increasing the volume, which they occupy in the velocity space. However,
the multiplication $dx^3\cdot dv^3$ should be conserved in accordance with the Liouville's theorem.
Thus, the space volume occupied by the neutrinos shrinks, and their density grows.

The galactic escape speed near the Solar System indubitably exceeds $525$~{km/s} \citep{escape} and
in principle may be much larger \citep{bt, suchkov}. If we accept $\vesc=600$~{km/s} and
$m_3=0.08$~{eV} (i.e., $\xi\simeq 630$~{km/s}), we obtain $b\simeq 1.15$. The obtained value of the
gravitational boosting of neutrino density are in good agreement with the N-body estimations of
this quantity \citep{desalas2017, zhang2018, mertsch2020}.

\subsection{The Laniakea Supercluster}

It follows from (\ref{23a15}) that the density boosting is defined by the gravitational potential
$\phi$ of the observer. The contribution of the gravitation of the Galaxy into the potential near
the Solar System is large, but probably not dominant. The Galaxy is surrounded by huge
extragalactic structures (like the Great Attractor), which are less dense, but much more massive,
and their gravitational potential can be at least comparable with the Galactic one. Unfortunately,
the influence of the nearest superclusters on the relic neutrino distribution cannot be calculated
precisely: the structure shape is rather complex, and numerical simulations seem to be the only
method. The properties of neutrinos differ significantly from the properties of the cold dark
matter, while even the single-component N-body simulations of the cold dark matter suffer from
essential numerical issues \citep{21, 17, 18}. Moreover, the superclusters are dominated by the
dark matter, and the observational determination of their masses and space structures are not very
precise, since the main component is invisible.

\begin{figure}
    \resizebox{1.0\hsize}{!}{\includegraphics{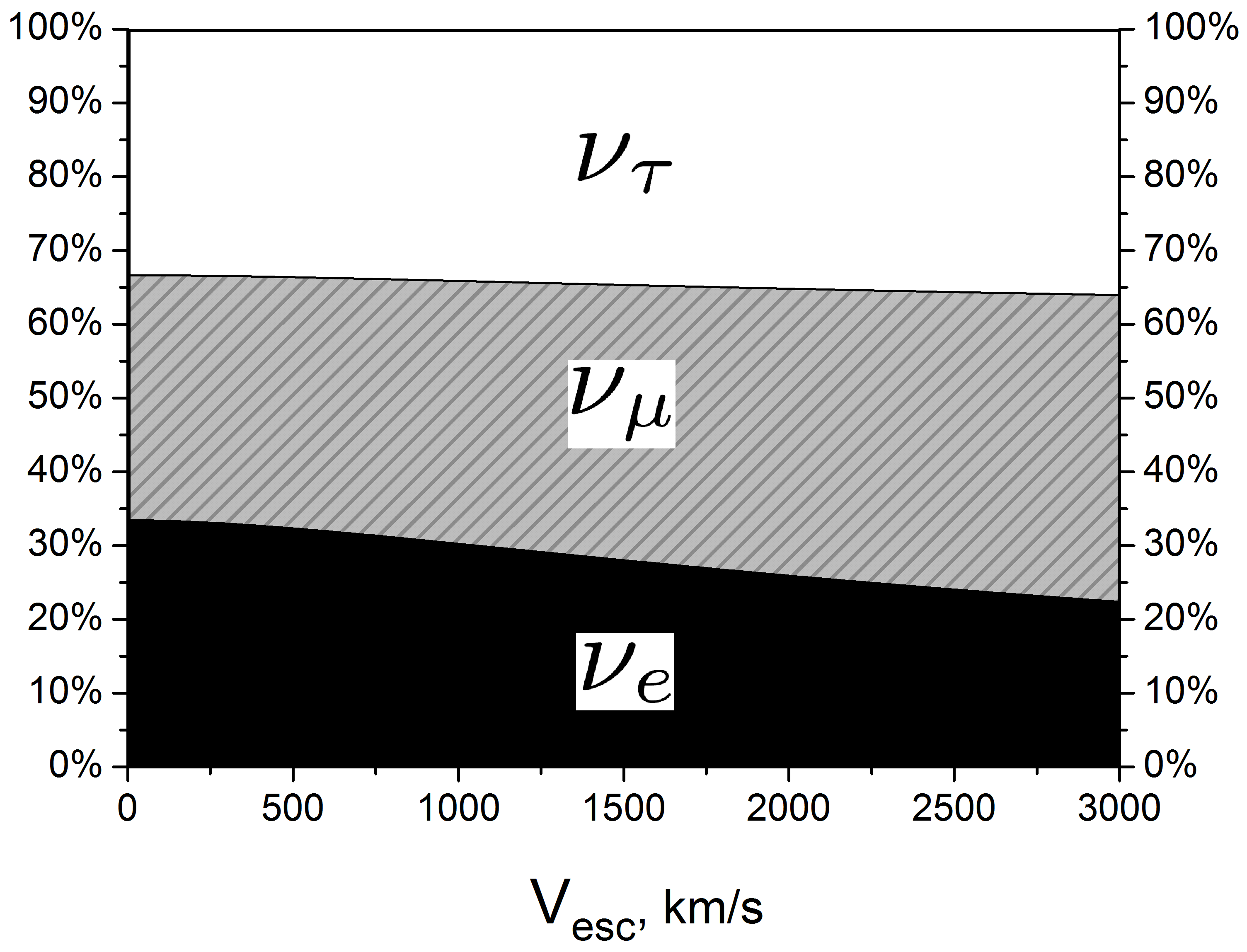}}
    \vspace{-0.0cm}
    \caption{The relic neutrino composition as a function of the total escape speed
$\vesc$. We accept $m_3=0.08$~{eV} and neglect the boosting of the two light neutrinos, $\nu_1$ and
$\nu_2$.} \label{23fig2}
\end{figure}

However, we may estimate the influence of the nearby objects with the help of
equation~(\ref{23a14}): we need just to estimate the total gravitational potential created by the
large scale structures, add it to the galactic potential, and substitute the obtained value
to~(\ref{23a14}). Let us justify this approach.

First of all, we need to specify that we use the comoving frame of reference (the standard
Friedmann coordinates). Then by a 'velocity' is meant a 'peculiar velocity', i.e., the velocity
with respect to the relic radiation field. This specification is not important in the case of the
Galaxy, but is essential for large structures.

Let us follow the derivation of~(\ref{23a14}). We may choose $r_2$ so large that the nearby galaxy
clusters are inside it. Then $f_2$ is defined by equation~(\ref{23a9}), and $\phi$ in~(\ref{23a14})
is defined not only by the gravitational field of the Galaxy, but by the nearby galaxy clusters as
well. Equation~(\ref{23a13}) in the derivation of~(\ref{23a14}) is not valid anymore, since the
system is not spherically symmetric. However, the final conclusion, $f_2=f_1$, does not depend on
the spherical symmetry and remains correct: it is just a consequence of the Liouville's theorem.

The distribution $f_2$ at the infinity depends only on the particle speed, and the change of the
particle speed depends only on the potential difference between $r_2$ and the observational point
$r_1$. In the case of the Galaxy, we may conclude on this ground that the distribution $f_1$ at the
observational point is also isotropic and depends only on the particle speed (which directly leads
to distribution~(\ref{23a14})). However, contrary to the case of the Galaxy, we cannot be sure that
$f_1$ is isotropic. The principle difference is that the gravitational field of the present Galaxy
is rather stationary, while a supercluster significantly evolves in the time necessary for a relic
neutrino to cross it. As a result, the potential $\phi$ changes as the neutrino moves from the
sphere $r_2$ to the observer, and the final neutrino speed depends on the neutrino trajectory. So
equation~(\ref{23a14}) is much less accurate in this case, than in the case of the Galaxy. However,
equation~(\ref{23a14}) is quite applicable as an approximation, since the logic of its derivation
remains valid: the gravitational field accelerates the neutrinos, decreasing the space value $dx^3$
that they occupy.

Thus, we need to find the total gravitational potential $\phi$ near the Solar System. Today this
task cannot be solved precisely, but we need only a rough estimation of this quantity. The
gravitational potential is additive, and we just sum the potentials from various massive objects,
neglecting the contributions that are significantly smaller then the galactic one
$\sqrt{2\phi_{MW}}\simeq 600$~{km/s}. It is easy to see that the potentials of the other members of
the Local Group or of the Council of Giants are negligible with respect to $\phi_{MW}$. However,
there are at least two objects that should undoubtedly be taken into account.

First of them is the Virgo cluster. Its mass inside $2.2$~{Mpc} from the center is estimated as
$M\simeq 1.2\cdot 10^{15} M_\odot$ \citep{virgomass}, and the distance to its center is
$d=16.5$~{Mpc} \citep{virgodistance}. Assuming that all the Virgo mass lies inside $2.2$~{Mpc}, we
estimate the potential created by the Virgo cluster near the Local Group as $\phi_V=GM/d$, or
$\vesc=\sqrt{2\phi_V}\simeq 790$~{km/s}.

The Local Group, as well as the Virgo cluster, are parts of the Laniakea Supercluster. The mass
distribution of this Supercluster is poorly known: some of its area projects on the Zone of
Avoidance, making the objects there essentially undetectable. As the second massive object we
consider the Great Attractor, a gravitational anomaly, which is supposed to be a massive galaxy
cluster obscured by the Milky Way disk. Its mass and the distance to this object are
estimated\footnote{These values are questionable now, since there is a debate about the Great
Attractor mass distribution. It is quite possible that the gravitational anomaly, which was earlier
considered as a single cluster, is a result of a joint action of several massive clusters (see, for
instance, \protect\citep{Kraan2016}). However, we need just a simple and rough estimation of the
gravitational potential, and we use these values for simplicity sake and to avoid the complex and
not quite clear question of the Laniakea mass distribution near the Great Attractor.} as $M\sim
5\cdot 10^{17} M_\odot$ and $d\simeq 70$~{Mpc} \citep{greatattractor}. It is important to underline
that the total mass inside $70$~{Mpc} from the Great Attractor center far exceeds this value.
However, calculating the potential, we should take into account only the overdensities, i.e.,
'additional' masses with respect to the homogeneous Friedmann's universe. We obtain $\vesc =
\sqrt{2\phi_{GA}} = \sqrt{2 GM/d}\simeq 2500$~{km/s}.

The total gravitational potential $\phi$ near the Solar System can be obtained by summation
$\phi_{MW}+\phi_{V}+\phi_{GA}$. Since the last quantity is very uncertain, we may conclude that the
total $\vesc$ lies most probably between $\sqrt{2(\phi_{MW}+\phi_{V})}\simeq 1000$~{km/s} and
$3000$~{km/s}. For $\nu_3$ (if we accept $m_3=0.08$~{eV}) these values correspond to the boost
factors $b=1.34$ and $b=2.66$, respectively. The boost factors for the light neutrinos $\nu_1$ and
$\nu_2$ are much less definite, because their masses and characteristic speeds $\xi$ vary in a wide
range, but even if we take the biggest possible mass $0.02$~{eV} and the highest speed $\vesc =
3000$~{km/s}, we obtain $b=1.22$, i.e., the boosting factors for the light neutrinos are much lower
than for $\nu_3$, and we will neglect them in this letter. However, they also may be important, if
$\vesc$ is high.

Finally, a question appears: if the escape speed is so high, is it correct to neglect the relic
neutrino capturing by the above-mentioned structures? The answer is positive, if we discuss the
relic neutrinos near the Solar System. Indeed, the Virgo cluster potential $\phi_{V}$ is comparable
with $\phi_{MW}$, and we could see that the fraction of the captured neutrinos is negligible in
this case. The gravitational potential of the Laniakea Supercluster is much larger, but it is not a
gravitationally bound object. We take into account only the peculiar velocities in our
calculations, while the Laniakea Supercluster participates in the general Hubble-Lema\^\i tre
expansion of the Universe, and the Great Attractor, for instance, moves away from us with the speed
$5000$~{km/s}. No capturing is possible under these conditions.

A significant quantity of relic neutrinos should be captured in the central areas of rich galaxy
clusters (like the Virgo cluster), where the escape speed is very high, and the captured neutrinos
may even dominate. However, even if the experimentalists ever manage to detect the relic neutrinos,
it will hardly happen in the center of a rich galaxy cluster, which reduces the practical
importance of this fact.

\section{The flavor composition of relic neutrinos}

A detection of relic neutrinos is a longed-for, but a notoriously difficult task. The cross-section
of neutrino interactions increases with the neutrino energy $e$ as $e^2$, but the Earth is
transparent even for neutrinos with $e\sim 1$~{MeV}, and one may imagine, how small the
cross-section is for the relic ones. Despite that, the neutrino community is offering new ideas
about possible ways of the relic neutrino detection, and some of them are even close to the
practical implementation (see, for instance, \citep{ptolemy}).

All the possible experimental methods have one common feature: they are based on the neutrino
flavors. The experimentalists observe $\nu_e$, $\nu_\mu$, $\nu_\tau$, and not the mass states,
$\nu_1$, $\nu_2$, and $\nu_3$. As we could see, the density of $\nu_3$ is significantly boosted by
the cosmic structures ($b=1.34-2.66$), while the density of the two others is boosted much less, if
any. So the fraction of $\nu_3$ is significantly boosted in the relic neutrinos. But $\nu_3$ is
almost free of the electron neutrino (only $\sim 2\%$ of $\nu_e$) and contains more muon component
than the tau one.

Being generated, the relic neutrinos have equal fractions of $\nu_e$, $\nu_\mu$, $\nu_\tau$, and we
can see that the gravitation field of the cosmic structures changes this composition.
Figure~\ref{23fig2} shows the relic neutrino composition as a function of the total escape speed
$\vesc$. We accepted $m_3=0.08$~{eV} and neglected the boosting of the two light neutrinos, $\nu_1$
and $\nu_2$. One may see that the fraction of $\nu_\tau$ remains almost constant (this is a result
of the fact that the fraction of $\nu_\tau$ in $\nu_3$ is $41\%\simeq 1/3$), while the fraction of
$\nu_e$ significantly decreases, and the fraction of $\nu_\mu$ significantly increases with the
$\vesc$ growth. If $\vesc = 1000$~{km/s}, the relic neutrinos contain $\sim 30\%$ of $\nu_e$, $\sim
36\%$ of $\nu_\mu$, and $\sim 34\%$ of $\nu_\tau$; if $\vesc = 3000$~{km/s}, the relic neutrinos
contain $\sim 22\%$ of $\nu_e$, $\sim 42\%$ of $\nu_\mu$, and $\sim 36\%$ of $\nu_\tau$.

Since the mass of a particle is always equal to the mass of the antiparticle, the densities of
relic neutrinos of any mass or flavor and the densities of corresponding antineutrinos remain
exactly equal.

Unfortunately, it is the rarest electron neutrino $\nu_e$ that is the most accessible to
observation. To comfort the experimentalists, we underline that the decreasing of the $\nu_e$
fraction is not a result of the $\nu_e$ density decreasing (the low limit on this density is given
by equation~(\ref{23a9}) for the undisturbed relic background), being a result of higher densities
of the two others (mainly of $\nu_\mu$).

In conclusion, let us discuss the highly unlikely case of the inverted hierarchy of neutrinos. Then
the situation is the opposite: $\nu_1$ and $\nu_2$ are heavy, $m_2\gtrsim m_1\sim 0.05$~{eV}, and
$\nu_3$ is much lighter. As a result, the densities of $\nu_1$ and $\nu_2$ are boosted by the
gravitational effects, and $\nu_2$ is the most abundant. The effect on the flavor composition is
also the opposite in this case: the relic neutrinos contain more electron neutrinos and less muon
neutrinos. Unfortunately, this type of hierarchy is now almost excluded experimentally.

\section{Acknowledgements}
Our pleasant duty is to thank Prof. Vadim Naumov (BLTP JINR) for his kind and valuable help in
calculation of the neutrino parameters. We would like to thank the Heisenberg-Landau Program, BLTP
JINR, for the financial support of this work. This research is supported by the Munich Institute
for Astro- and Particle Physics (MIAPP) of the DFG cluster of excellence "Origin and Structure of
the Universe".


\label{lastpage}
\end{document}